\documentclass[11pt]{article}
\usepackage[utf8]{inputenc}
\usepackage[T1]{fontenc}
\usepackage{lmodern}
\usepackage[margin=2.54cm]{geometry}
\usepackage{graphicx, subcaption}
\usepackage[flushleft]{threeparttable}
\usepackage{booktabs}
\usepackage{array}
\usepackage{caption}
\usepackage{natbib}
\usepackage[hidelinks]{hyperref}
\usepackage{xurl}
\usepackage{microtype}
\usepackage{amsmath} 
 
\title{\textbf{Could Underwater Data Centers Pose a Risk\\ to AI Treaty Verification?}}
\author{James Teague\thanks{Corresponding author: \texttt{J.Teague@sms.ed.ac.uk}} \and Ashmita Rajmohan  \and Yannick Muehlhaeuser}

\date
{}
 
\begin{document}
\maketitle

\begin{abstract}
\noindent
Proposals for international agreements that limit frontier AI development depend on verification, and a central challenge is detecting undeclared compute facilities used to evade restrictions. Underwater data centers (UDCs) have been suggested as one such evasion vector, but their feasibility at frontier scale and their detectability have not been seriously assessed. We examine current UDC deployments, evaluate construction and maintenance complexity relative to land-based facilities, and analyse the feasibility of a 100{,}000 H100-equivalent training run underwater. We find that power delivery and cooling are tractable, but interconnect and the hands-on maintenance that large training runs require are severe obstacles — surmountable only by a well-resourced state actor accepting large cost and schedule penalties, and only where concealment, rather than efficiency, is the objective. We then assess detectability through thermal, acoustic, optical and synthetic-aperture-radar (SAR) surveillance. Thermal detection of an operational pod is unlikely outside shallow, calm water; acoustic detection is marginally more effective, but faces limitations in attribution; and optical/SAR monitoring is most powerful during construction and maintenance, when the pressure-vessel fabrication base and the cable-laying fleet create distinctive signatures for AIS-tracking. We conclude that UDCs are a comparatively unlikely evasion route relative to underground or industrially disguised land-based facilities, but the residual risk is non-zero and warrants operationalising the detection modalities discussed.
\end{abstract}

\vfill
\vspace{6pt}

\subsubsection*{Acknowledgements}
{\small 
This work was done as part of the Orion AI Governance Initiative by Arcadia Impact. We thank Joe Hardie for general support and comments. Further input was provided by Christina Krawec, Joshua Turner, Mauricio Baker, Ben Harack, Mehmet Sencan, James Petrie, Robi Rahman and Aaron Scher.}

\newpage

\section{Introduction}

Most proposals for international agreements that place restrictions on frontier AI development require robust verification mechanisms to monitor compliance \citep{scher2025}\footnote{Likewise, even without international agreements, deterrence regimes depend upon the ability to locate a rival's compute \citep{hendrycks2025}}. A critical challenge for such agreements is detecting undeclared or ``dark compute'' facilities that states might deploy to evade such restrictions. Underwater data centers (UDCs) have been proposed as a possible evasion vector. However, the technical feasibility of deploying UDCs at the scale required for training frontier models remains unclear, and their detectability has not been seriously investigated. We examine the state of current UDC deployments, evaluate construction complexity relative to traditional data centers, and investigate their feasibility for large-scale AI training. We then assess the detectability of subsea clusters via thermal, acoustic, and visual surveillance.

\section{How Underwater Data Centers Work}

UDCs are sealed capsules of computing infrastructure deployed on the seabed, using cold ocean water for cooling instead of conventional systems \citep{semicooling}. This section covers their cooling methods, physical design, operational reliability, and current deployments.
 
\subsection{Cooling}
UDCs use two primary cooling methods. In the first, cold seawater is piped through heat exchangers positioned behind the server racks \citep{cutler2017}. In the second, servers are immersed in a dielectric fluid that absorbs heat directly from the chips, which is then dissipated to the surrounding seawater \citep{hu2022, muneeshwaran2023}. Both approaches eliminate the chillers, cooling towers, and HVAC systems that dominate the energy budgets of land-based, air-cooled facilities, where HVAC has historically accounted for $\sim$\,25--40 percent of electricity consumption \citep{shehabi2016}. 

Importantly, existing UDCs have been deployed at 9--36\,m depths in a seasonally-mixed surface layer (Table~\ref{tab:specs}), rather than the stable, near-freezing water that deep seawater cooling relies on ($\geq$\,100\,m). Even at these shallow depths the thermal advantage is substantial: Project Natick held cooling overhead to 3 percent of total facility power at 11\,m \citep{cutler2017}. Greater depths would offer a colder and more stable heat sink, but at the cost of greater structural and installation demands \citep{kumar2025}.
 
\subsection{Physical Design}
The basic design is a sealed, pressure-resistant steel cylinder for multi-year unmanned operation, sometimes referred to as ``lights-out'' deployment \citep{natick}. Each capsule is connected to shore by a submarine cable bundle that carries electrical power and fibre-optic data links: functioning as an extension of a land-based network \citep{microsoft2020}. Deployment and retrieval are handled by crane barges or similar marine vessels, which lower the capsules and anchor them to the seabed, and reverse the process when a unit reaches end of life or requires servicing \citep{dcd2025natick}.

Two main design families have emerged, each pairing a pressure strategy with a corresponding cooling approach. Sealed capsules, as used in Project Natick, maintain one atmosphere of internal pressure and are filled with dry nitrogen, which eliminates the oxygen and humidity that drive corrosion in conventional facilities; heat is moved from the servers to the ocean via fans and seawater piped through internal heat exchangers \citep{natick}. Pressure-equalised pods, as proposed by Subsea Cloud, match internal pressure to the surrounding ocean. Because gases compress at depth, pressure equalisation requires filling the enclosure with a nearly incompressible dielectric fluid, which doubles as the immersion coolant: servers sit directly in the fluid, and heat transfers through the container walls to the surrounding seawater. In principle, this reduces structural requirements, since the walls no longer need to withstand the full pressure differential \citep{register2022, dcdsubsea2022}.
 
\subsection{Reliability}
Over its two-year deployment, Natick recorded less than one percent of servers failing, approximately one eighth the rate of equivalent land-based facilities \citep{microsoft2020}. This improvement has been attributed to the inert nitrogen atmosphere inside the capsule and the absence of human-induced disruptions like accidental bumps, cable pulls, and maintenance errors that are common in staffed facilities \citep{natick}.
 
However, because the capsules are sealed and inaccessible during operation, individual server failures could not be repaired in situ like land-based data centers, where full repairs can be completed within hours and training continuity can be restored within minutes. Instead, workloads are redistributed across remaining healthy nodes, and failed hardware stays offline until the capsule is retrieved at the end of its operational cycle \citep{microsoft2020}. For sustained deployments spanning two years or longer, these individual failures could compound, likely rendering the long-term performance of UDCs progressively worse than equivalently provisioned land-based clusters; at least until a reliable mechanism for conducting repairs whilst pods remain operational is found.
 
\subsection{Current and Past Deployments}
Microsoft's Project Natick was the first UDC project. Phase 1 deployed a single rack off the California coast from August to December 2015. Phase 2, a full-scale 864-server module, operated on the seabed off Orkney from June 2018 to July 2020. Natick was formally discontinued in 2024 \citep{dcd2024natickend}, with the difficulty of servicing, upgrading and replacing sealed hardware widely suggested as the reason --- a concern that the project lead had earlier dismissed \citep{dcd2025natick}.\footnote{See Appendix B Figure \ref{fig:Natick1} for an image of the Phase 1 Natick pod.} 

\begin{figure}[!htbp]
\centering
    \begin{subfigure} {.4\textwidth} 
    \centering 
    \includegraphics[width=1\linewidth]{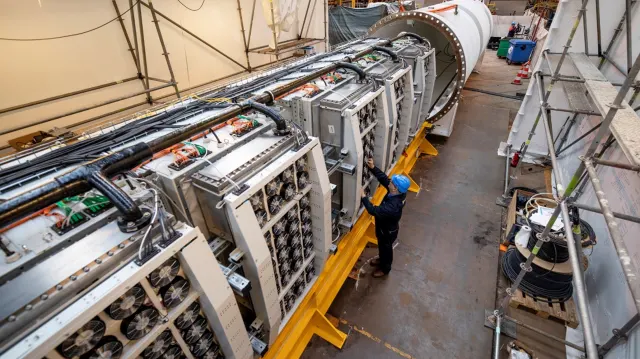}
    \caption {Construction of the Phase 2 pod in France.}
    \end{subfigure}
    \hspace{0.03\textwidth}
     \begin{subfigure} {.4\textwidth}
         \centering 
    \includegraphics[width=.88\linewidth]{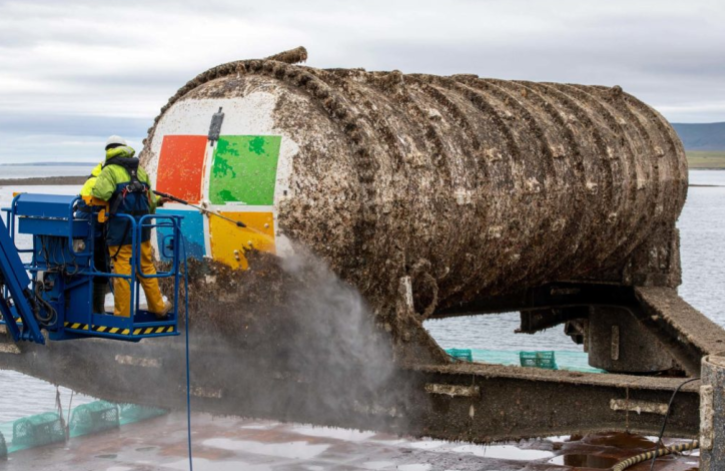}
    \caption{The recovered pod after 2 years.}
\end{subfigure}
\caption{Project Natick Phase 2 -- Northern Isles. (Images courtesy of Microsoft.)}
\end{figure}
\begin{figure}[!htbp]
    \centering
    \includegraphics[width=0.4\linewidth]{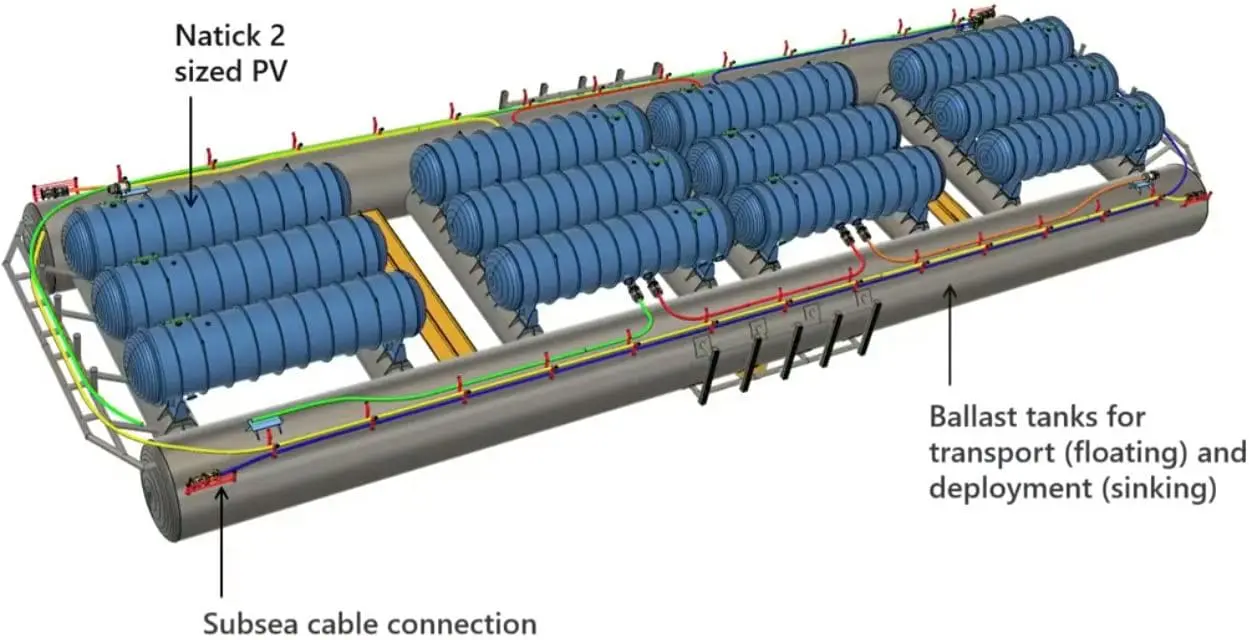}
    \caption{A depiction of Natick Phase 3. (Image courtesy of Microsoft).}
    \label{fig:natick3}
\end{figure}

\newpage
Highlander (via subsidiary HiCloud) is the only company currently operating commercial UDCs. Its Hainan cluster has grown incrementally toward a long-term target of 100 modules: from initial tests in 2021 \citep{judge2021highlanderudc}, to a first commercial module deployed in 2023 \citep{tomshainan2023}, and a second commercial module in February 2025, adding roughly 400 servers \citep{ecns2025}. Its Shanghai facility completed Phase 1 construction in October 2025 and entered operation in May 2026 \citep{offshorewindbiz2026}, making it the first UDC directly linked to an offshore wind farm. Phase 1 was operating at 2.3\,MW in May 2026, against a 24\,MW power capacity in the full build-out. Importantly, the Shanghai facility is not completely submerged, and will therefore not be our focus for the remainder of this paper.

\begin{figure}[!htbp]
\centering
    \begin{subfigure} {.43\textwidth}
    \centering 
\includegraphics[width=.9\linewidth]{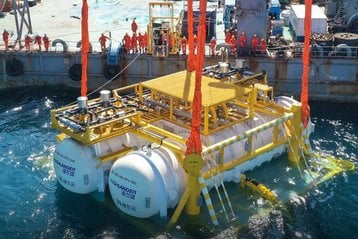}
    \caption {Highlander Module 2 is deployed in Hainan in February 2025.}
    \end{subfigure}%
    \hspace{0.03\textwidth}
     \begin{subfigure} {.43\textwidth}
         \centering
     \includegraphics[width=1\linewidth]{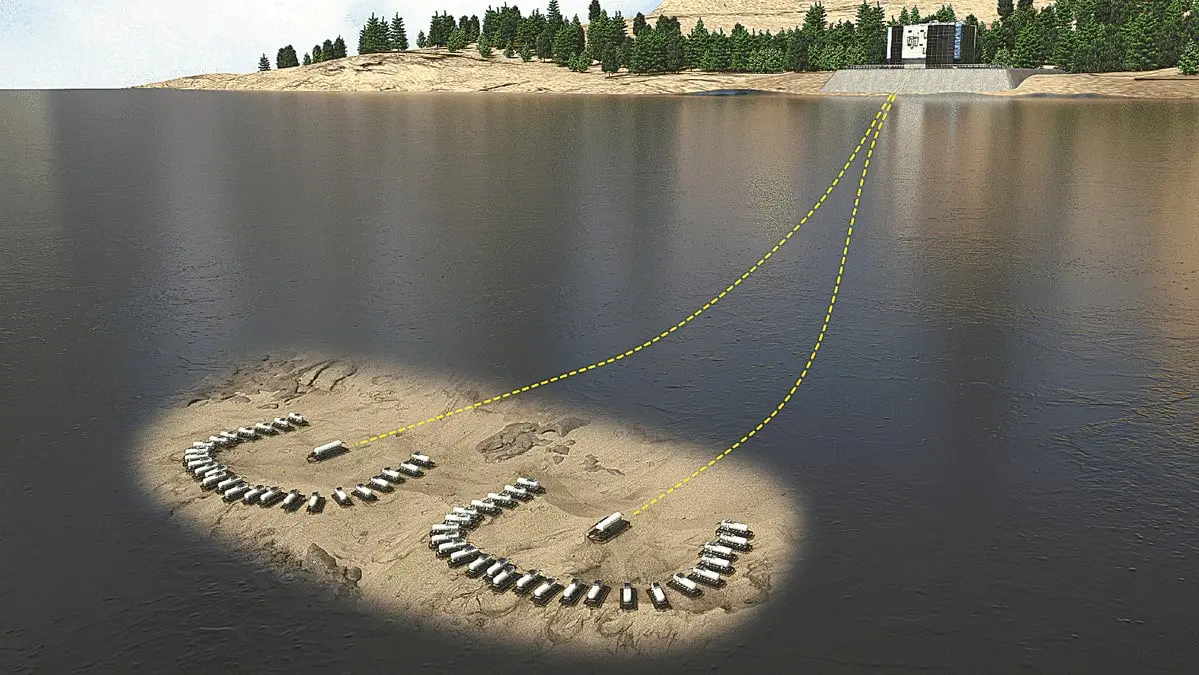}
    \caption{HiCloud visualisation of the pods on the ocean floor, connected to land by cabling.}
\end{subfigure}
\caption{Highlander: Hainan deployment (Images courtesy of China Daily).}
\end{figure}
\begin{figure}[!htbp]
\centering
\begin{subfigure} {.43\textwidth}
    \centering 
\includegraphics[width=.6\linewidth]{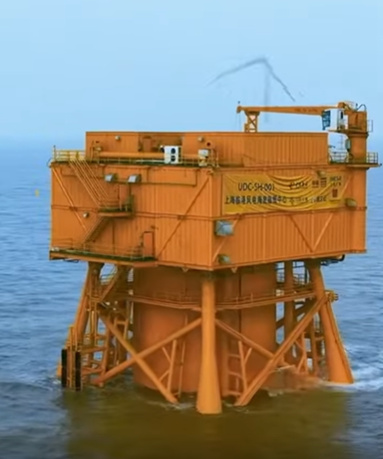}
    \caption {The 2.3\,MW module operating off the coast of the Lingang Special Area Trade Zone.}
    \end{subfigure}%
\begin{subfigure} {.43\textwidth}
    \centering 
\includegraphics[width=.9\linewidth]{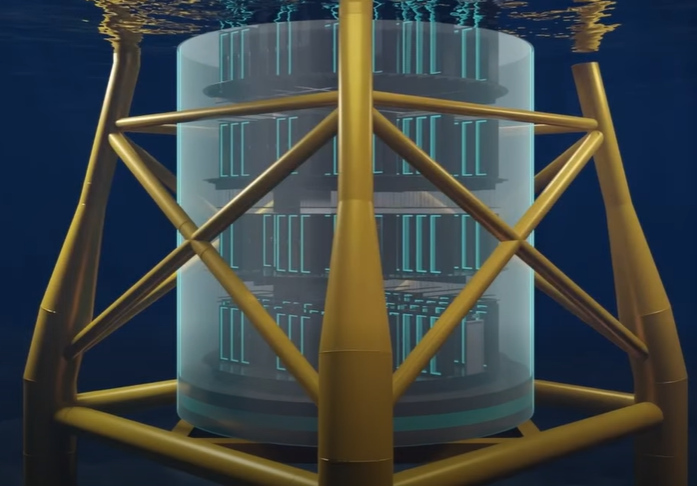}
    \caption {Visualisation of the server tiers in the wind-powered module.}
    \end{subfigure}%
    \caption {Highlander: Shanghai deployment (Images courtesy of China Daily).}
    \end{figure}
    
Subsea Cloud, a US based company, has announced Jules Verne, a demonstration pod near Port Angeles, Washington, and Project OTTO, a rolling trial off south-west Norway whose first phase was scheduled for 2024 \citep{dcdsubsea2022}. The Jules Verne design is a 20-foot container that uses dielectric immersion-cooling: holding 16 racks and $\sim$\,800 servers at 9\,m depth \citep{register2022}, drawing $\sim$\,1\,MW (advertised as 0.5--1.8\,MW \citep{subseaabout}). Importantly, we could not find independent verification of any operating UDCs from Subsea Cloud.\footnote{See Appendix B Figure \ref{fig:SubseaVerne} for a rendering of the Jules Verne pod.}
 
Network\-Ocean, a Y Combinator-backed startup, has announced a 0.5\,MW test capsule for San Francisco Bay but has faced regulatory obstacles, with two California agencies flagging a lack of required permits \citep{dcdnetworkocean}. It plans to submerge the capsule a few metres below the surface for an hour, claiming the capsule would eliminate on-site water consumption and reduce power usage up to 30 percent.\footnote{See Appendix B Figure \ref{fig:Network} for a picture of Network\-Ocean's 0.5\,MW pod under construction.}

\clearpage
\begin{table}[ht!]
\centering
\begin{threeparttable}
\caption{Specifications of selected UDC deployments. Figures are drawn from vendor disclosures and press reporting, so many are not independently verified.}
\label{tab:specs}
\footnotesize
\setlength{\tabcolsep}{1.8pt}
\renewcommand{\arraystretch}{1.25}
\begin{tabular}{>{\raggedright\arraybackslash}p{1.7cm}
                >{\raggedright\arraybackslash}p{2.0cm}
                >{\raggedright\arraybackslash}p{2.1cm}
                >{\raggedright\arraybackslash}p{2.1cm}
                >{\raggedright\arraybackslash}p{2.1cm}
                >{\raggedright\arraybackslash}p{2.1cm}
                >{\raggedright\arraybackslash}p{2.1cm}}
\toprule
\textbf{Spec} & \textbf{Natick Ph1 (2015)}\tnote{a} & \textbf{Natick Ph2 (2018--20)}\tnote{b} & \textbf{High\-lander Hainan (2025)}\tnote{c} & \textbf{High\-lander Shanghai (2026)}\tnote{d} & \textbf{Subsea Cloud Jules Verne}\tnote{e} & \textbf{Network\-Ocean}\tnote{f} \\
\midrule
Dimensions & $\sim$3\,m $\times$ 2.1\,m & 12.2\,m $\times$ 2.8\,m & $\sim$18\,m $\times$ 3.6\,m & 4 server tiers, $\sim$160\,m$^2$ each; 1{,}950\,t total & $\sim$20\,ft $\times$ 8.5\,ft (6\,m) container & Unknown (estimated 2--3\,m $\times$ 6--9\,m)  \\
Depth & $\sim$11\,m & $\sim$36\,m & $\sim$35\,m & $\sim$10--35\,m & $\sim$9\,m & ``Just under surface'' \\
Servers / racks per pod & 1 rack, $\sim$\,24 servers, 1 vessel & 12 racks, 864 servers, 1 vessel & $\sim$800 servers, 2 vessels (400/pod)& 192 racks, 1 vessel \newline(Phase 1) & 16 racks, $\sim$800 servers/pod & 1 test capsule planned \\
CPU / accelerator & Likely Azure Hardware \space(e.g. Xeon) & Xeon E5 v3/v4 + Arria 10 FPGAs; no GPUs. & Undisclosed & Undisclosed & Undisclosed & NVIDIA H100 (claims 2{,}048 reserved) \\
Total power & Low (test-scale) & 240\,kW & Unclear & Ph1: 2.3\,MW; 24\,MW full construction capacity & 1\,MW (0.5--1.8\,MW advertised) & 0.5--1\,MW test; barges to 200\,MW+ \\
Compute (FLOPS) & Estimated $\sim$\,10s of TFLOPS (low) & Estimated $\sim$\,100s of TFLOPS (low) & No benchmark; 2 exaFLOPS planned in 100-module buildout & Undisclosed & Undisclosed & Planned barge $\sim$2{,}026 PFLOPS FP16 (2{,}048 $\times$ H100) \\
Power source & Shore grid & Shore grid & Shore grid & Offshore wind via submarine cable ($>$95\%); grid backup ($<$5\%) & Not specified (Likely shore grid) & Not specified (Likely shore grid) \\
\bottomrule
\end{tabular}

\begin{tablenotes}[flushleft]\footnotesize
\item[a] Microsoft Project Natick Phase 1 (\emph{Leona Philpot}), deployed off San Luis Obispo, California, August--December 2015. Deployment depth and the single-rack configuration are reported by the Natick team \citep{cutler2017}; the capsule measured 10\,ft $\times$ 7\,ft ($\sim$3.05\,m $\times$ 2.13\,m) \citep{engnatick2015}.
\item[b] Microsoft Project Natick Phase 2, deployed off Orkney, June 2018 -- July 2020. Capsule dimensions, deployment depth, rack and server counts, hardware configuration and power draw \citep{natick}.
\item[c] Highlander/HiCloud, Hainan. Capsule dimensions \citep{ecns2025}; deployment depth, capsule weight and the 100-module target \citep{rmrb2024, yicai}; servers/pod and second-module details \citep{devtelecoms}.
\item[d] Highlander/HiCloud, Shanghai Lingang, Phase 1. Structure, tier layout, weight, depth and as-built rack count \citep{starmarket2026}; commissioning \citep{motgov2026, offshorewindbiz2026}; power figures \citep{tomswind, txnews2025shanghai, ieshanghai}. 
\item[e] Subsea Cloud, \emph{Jules Verne} demonstration pod, Port Angeles, Washington. Container dimensions, rack and server counts, and capacity \citep{register2022}; deployment depth \citep{dcdsubsea2022}. The vendor's current product page gives a wider 0.5--1.8\,MW range across the module line \citep{subseaabout}.
\item[f] NetworkOcean, announced test capsule for San Francisco Bay; not deployed at time of writing. Depth and permitting status \citep{dcdnetworkocean}; test capsule plan, the 2{,}048-GPU reservation and power target \citep{ycnetworkocean, dcdnetworkocean}; H100 FP16 throughput used for barge projection \citep{nvidiah100}.
\end{tablenotes}
\end{threeparttable}
\end{table}

\section{Construction and Maintenance Complexity}

Building a UDC draws on a different industrial base than building a data center on land. A conventional facility requires concrete, structural steel, and HVAC contractors: industries with tens of thousands of qualified firms worldwide. An underwater facility requires pressure-vessel fabrication, submarine cable installation, and marine engineering capability, and faces severe limitations in maintenance, hardware refresh, and potential vulnerability to sabotage.
 
\textbf{Hull fabrication.} Currently, the core structural element of a UDC is a sealed, corrosion-resistant steel pressure vessel. Approximately 14--17 countries currently possess the capability to design and build full-sized submarines \citep{sutton}\footnote{Major yards are in China (CSSC/CSIC, including Wuchang), France (Naval Group), South Korea (Hanwha Ocean, HD Hyundai), Japan (Mitsubishi, Kawasaki), Germany (ThyssenKrupp Marine Systems), Sweden (Saab Kockums), and several others including India, Italy, and Turkey.}, making pressure-vessel fabrication a strong supply-chain chokepoint: the activity is concentrated in a small number of known, large shipyards and is inherently visible to satellite observation. However, this bottleneck may not be permanent. With significant investment in scaling pressure-vessel fabrication, the capsules may become factory-built modular units produced in serial production across multiple yards, eroding both the cost premium and the monitoring advantage of supply-chain concentration.
 
\textbf{Submarine cabling.} Every UDC requires at least one submarine cable bundle carrying both electrical power and fibre-optic data links from shore. Installing and maintaining these cables requires specialised cable-laying vessels, of which it is estimated there are around 60 globally \citep{dcdcable}.\footnote{Not all 60 are equivalent: many are smaller repair vessels or fibre-only ships. Power cable installation, which is what an underwater data center requires, needs larger vessels with higher-capacity carousels and tensioners, further narrowing the available fleet to perhaps 20--30 ships.} The major operators are a known and small set of companies\footnote{Prysmian, SubCom, Nexans, NEC, Global Marine, Jan De Nul and SBSS.}, and their vessels, as commercial shipping, carry AIS transponders and are therefore trackable in principle.
 
\textbf{Marine operations.} Activities such as cable laying, seabed preparation, pod deployment via crane barge, and umbilical connection account for a substantial share of total UDC project cost. This is plausibly the single largest cost premium over land-based construction, where site preparation and utility connection, while slow, draw on abundant and inexpensive civil engineering resources.
 
\textbf{Maintenance and Hardware refresh.} In the current generation of UDCs, hardware failures such as a dead GPU or a failed network switch cannot be repaired until the entire pod is retrieved. Subsea Cloud claims retrieval in 4--16 hours for its lighter, pressure-equalised pods \citep{register2022}.\footnote{This is an unverified vendor figure, reported at the company's 2022 announcement and covering transit to the site as well as the recovery itself; it has not been independently demonstrated at scale.} Similarly, the hardware refresh cycle could also be a significant bottleneck for UDC economics. GPU generations evolve every 12--18 months, and sealed pods designed for 3--5 year deployment cycles cannot accommodate mid-cycle upgrades.
 
\textbf{Sabotage.} A final point is the vulnerability to physical sabotage. Off-the-shelf underwater speakers, tuned to a hard-drive resonant frequency ($\sim$5\,kHz, within a measured range of $2.0$--$8.9$\,kHz), can displace drive read/write heads far enough to degrade RAID throughput and force nodes out of a distributed filesystem within minutes \citep{aquasonic2024}. This sits alongside other kinetic forms of sabotage. The 2022 Nord Stream attacks, which Denmark, Germany and Sweden jointly confirmed to the UN Security Council were caused by deliberate explosions \citep{nordstream}, demonstrate both the feasibility of striking state-scale subsea infrastructure and the difficulty of attributing responsibility afterwards.

\section{Feasibility of Frontier AI Training}

Power delivery and cooling for such a training run appear quite feasible. 100k H100s draw $\sim$70\,MW at the GPUs alone \citep{nvidiah100}, but server overhead (CPUs, NICs, PSUs) and switch fabric push this toward $\sim$150\,MW \citep{semianalysis100k}. A frontier cluster in the relevant 3--5 year window may also require denser hardware (H200s or B200s), suggesting 200\,MW as a plausible high-end. Submarine HVDC power cables routinely carry GW-scale loads \citep{subcable}, so delivering 150--200\,MW via a dedicated submarine power cable is well within proven engineering. Cooling is where UDCs have an advantage over land-based clusters, so it is unlikely that this would constitute a bottleneck.
 
Interconnect is a plausible limitation. For one, the network fabric itself is physically massive. Beyond the GPU racks, a 100k H100-equivalent cluster also needs hundreds of leaf switches and dozens of spine switches, each consuming tens of kilowatts, each requiring precisely structured cabling to every other switch. In a reference 8{,}192-GPU design, 256 leaf switches connect to 36 spine switches housed in dedicated spine racks \citep{drivenets}. This likely increases the required power delivery (accounted for above), and more importantly, means substantial additional engineering effort beyond what UDCs are currently designed for. While it may not be economical, this appears solvable for a motivated actor.
 
Cable distance is another aspect of networking that could constitute a bottleneck. Between racks, direct-attach copper works up to about 3\,m, with active optical cables and multimode fibre extending reach to roughly 50\,m for the leaf layer \citep{semianalysis100k}. This means the physical layout must be extremely dense and precisely arranged, which likely implies additional engineering effort and risk of malfunction.
 
The most significant obstacle from the networking side appears to be the required maintenance. Common maintenance tasks such as unplugging and reseating an InfiniBand transceiver are fundamentally hands-on operations \citep{semianeocloud2024}.\footnote{The most common operational issues in GPU clusters are ``flapping IB transceivers, GPUs falling off the bus, GPU HBM errors, and silent data corruption'', and resolution often requires ``unplugging and plugging back in the InfiniBand transceiver or cleaning the dust off of fiber cables''.} Meta's 405B training run for Llama 3 provides the most detailed public data point: over 54 days on 16{,}384 H100 GPUs, the cluster experienced 419 unexpected interruptions (approximately one every three hours) \citep{meta2024llama}. Optical failures, arguably the most severe reliability problem, also do not get reduced by a less corrosive atmosphere the way chip failures themselves are. This means that conducting a large training run in a UDC will either require very difficult underwater maintenance or massive redundancy to route connections around points of failure.
 
Overall, interconnect is where the constraints are most severe --- though not because the fabric itself is unbuildable. The switching layer, the cabling density it demands and the additional power it draws are all tractable for an actor willing to absorb the cost. The binding constraint is that this fabric requires continual hands-on maintenance: reseating transceivers, cleaning optics, and swapping failed links. A state actor could substitute capital for access, through heavy link redundancy and overprovisioned spare capacity, but even this would constitute a severe scheduling penalty. Whether UDCs constitute a viable dark-compute pathway therefore turns on the probability of detection.

\section{Detectability and Agreement-Evasion Feasibility}\label{sec:detect}

The detection modalities most applicable to operational UDCs are thermal infrared (TIR) imaging and acoustic surveillance, whilst optical and synthetic aperture radar (SAR) imaging are capable of detecting the logistics and infrastructure activity associated with building and installing a UDC cluster. This section will show that the effectiveness of TIR imaging is likely negligible given the effects of heat dispersion in coastal waters; acoustic detection is physically plausible but faces significant operational bottlenecks; and optical and SAR monitoring are most powerful during the deployment window.
\subsection{Thermal Detection}
A frontier-scale UDC cluster drawing $\sim$150--200\,MW would continuously reject waste heat into the surrounding water column through seawater heat exchangers or immersion-fluid systems. The relevant question is whether this produces a detectable sea-surface-temperature (SST) anomaly. At the $\sim$35\,m deployment depths of current commercial UDCs, and given that frontier-scale training would likely require an area on the order of $\sim$1.7--9 hectares~(Appendix~\ref{app:footprint}), detection is unlikely in the majority of open coastal conditions. Waste heat expelled at this depth would need to rise through a coastal water column before reaching the surface, where current velocity usually sits well above 0.17\,m/s, implying an SST change below the standard deviation of natural variability ($\pm$0.3--0.5$^\circ$C in lower-variability ocean basins \citep{deser2010}).~(Appendix~\ref{app:velocity})

Conventional TIR imaging satellites, such as Landsat-8, offer coarse resolution (100\,m) relative to the scale of a UDC's likely thermal signature, and are often limited by cloud-cover, creating persistent monitoring gaps. Whilst \citet{ahn2006} demonstrated that satellite thermal-infrared imagery (NOAA AVHRR and Landsat) could track the warm-water plume from Korea's $\sim$5{,}900\,MW Younggwang nuclear plant up to 100\,km from the outfall, this discharge enters the sea from a concentrated point at the surface rather than being dispersed at depth. A UDC's smaller, deeper discharge would not produce a comparable surface signature. A newer commercial thermal satellite does not close this gap either: SatVu's HotSat-2 offers day-and-night imaging at 3.5\,m resolution, yet its imager operates in the mid-wave infrared (approximately 3.7--5\,$\mu$m) rather than the longwave thermal-infrared band \citep{satvu}. Absolute sea-surface-temperature retrieval requires the paired thermal-infrared channels near 11 and 12\,$\mu$m used by sensors such as VIIRS, MODIS and Sentinel-3 SLSTR, whose brightness-temperature difference corrects for atmospheric water-vapour absorption, and have much coarser resolution \citep{minnett2019}. 

The three key variables affecting detectability are therefore depth, dispersion and location. Deeper and more widely dispersed pods will reduce the thermal signal at the sea surface, as discharge rises through a more stratified water column and over a greater volume. The additional costs and engineering constraints of deeper and more dispersed pods can be offset by building UDCs in waters with increased current velocity, significantly reducing the likelihood of a detectable SST anomaly from shallower or denser clusters. This effectively rules out UDC construction in calmer waters~(Appendix~\ref{app:temprise}) (e.g.\ coves, sheltered inlets, or inland waters such as lakes), where covert construction and operation may initially appear simpler. Operators seeking to evade thermal detection would therefore have strong incentives to deploy as deep as is structurally and economically feasible, as distributed as interconnect constraints allow, and in waters as turbulent as a hull can feasibly withstand without compromising efficiency.
 
\subsection{Acoustic Detection}

Both UDC design families generate an acoustic signature, but only one source is common to both: power conversion. A cluster fed by a submarine HVAC link steps voltage down locally, and the transformers radiate a stable tone at twice the supply frequency, 100 or 120\,Hz,
with harmonics. An order-of-magnitude scaling estimate places the total signature at roughly 127--147\,dB re 1\,$\mu$Pa at 1\,m.~(Appendix~\ref{app:sourcelevel})\footnote{Sealed capsules (as demonstrated by Project Natick) move heat with fans, adding shaft-rate tones (roughly 20--150\,Hz) and blade-passing tones an order of magnitude higher. For simplicity, and following scepticism from reviewers that sealed capsules would be used for frontier training runs, we focus on the acoustic signature of power conversion equipment, which also covers dielectric immersion cooled UDCs. Additionally, whilst immersion pods have no fan noise, their internal sources sit in a fluid whose acoustic impedance is much closer to seawater than the gas-to-steel boundary of a sealed hull, so the power-conversion tone couples out more efficiently. We expect, in principle, for these features to place both design families at a similar level of acoustic signature.}
 
Unlike submarines, which are moving targets that use active noise-reduction engineering, UDCs are stationary, continuously operating industrial facilities, suggesting that they may be easier to track via traditional acoustic surveillance systems. However, the US Navy's primary detection mechanism for submarines is the SOSUS network \citep{sosus}: arrays of hydrophones positioned to exploit the SOFAR channel \citep{sofar}, whose axis at mid-latitudes lies at roughly 600--1{,}200\,m. The system is capable of passively detecting acoustic power of less than one watt at ranges of several hundred kilometres, but it depends on coupling into that deep waveguide. Current UDC deployments at 30--35\,m sit in shallow coastal waters, where propagation losses are substantially higher and bottom reverberation masks nearby targets.
 
Without the monitored state's cooperation on naval monitoring, acoustic detection of a frontier-scale UDC would therefore likely require the covert deployment of sensors. Existing civilian infrastructure, such as oceanographic moorings or research hydrophone arrays, could provide data without active sensor deployments, but coverage near specific coastal sites cannot be assumed and would require deliberate pre-positioning under scientific cover. A purpose-built sensor such as the AN/SSQ-53 DIFAR sonobuoy \citep{fassonobuoy} can operate for up to 8 hours by transmitting data back by radio link, covers the transformer frequency range and harmonics, and can provide directional bearings to a source, such that two sonobuoys can triangulate a contact. Given the source level estimated above, and the transmission losses characteristic of 35\,m underwater depths, a sonobuoy would need to be positioned within roughly 5--20\,km of the cluster to achieve detection (with the upper end of that range dependent on bottom type and sea state).~(Appendix~\ref{app:detrange})
 
The practical implication is that passive acoustic monitoring would likely require the deployment of sensors inside the boundary of the target state's 12 nautical mile (22\,km) territorial sea \citep{unclos}. They could be deployed from a covert submarine, lowered from a surface vessel under civilian cover, or dropped from altitudes up to 30{,}000 feet. However, each method carries significant risk: a state concealing a UDC would plausibly monitor the surrounding water and likely approach routes for intruding submarines and suspicious foreign objects, and track anomalous surface-vessel activity using satellite AIS and coastal radar. A further limit is that detection is not the same as attribution. The mains-frequency tone that dominates the UDC signature is radiated by every shore-connected alternating-current installation in a coastal region: ships' generators, offshore platforms, wind farms and pumping stations all produce energy in the same narrow band. Detecting acoustic energy at 100 or 120\,Hz therefore identifies very little on its own: discrimination would rest on the harmonic structure of the source, on its spatial stationarity, and on the absence of any declared installation at that position, none of which a single passive sensor establishes.

\begin{figure}[!htbp]
    \centering
    \includegraphics[width=0.3\linewidth]{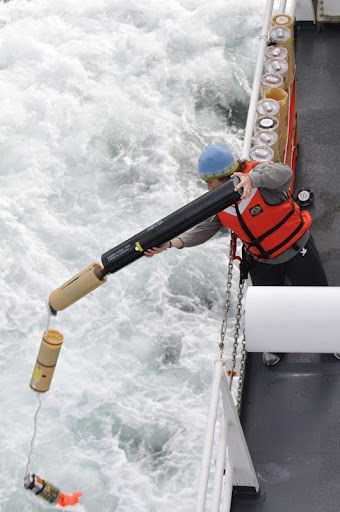}
    \caption{A sonobuoy is launched from the NOAA Ship Oscar Dyson.}
\label{fig:sonobuoy}
\end{figure}
 
 Confirming that a contact is a data center rather than ordinary marine infrastructure would therefore, in practice, likely require the use of active sonar. Small autonomous underwater vehicles (AUVs), such as the REMUS\,600 used for seabed survey work \citep{remus}, can image a strip of seabed with side-scan sonar in a single pass, so a cluster spanning $\sim$1.7--9 hectares could be covered in a modest number of transects. AUVs have many advantages over sonobuoys, such as producing a clear image rather than an inferred contact, and much greater efficiency in area covered over time. Against an alert host, however, this is not reliably stealthy, since active sonar pings the seabed, the vehicle's noise is highly detectable by purpose-built intruder sonar \citep{navylookout}, and its launch and recovery would expose deployers to the same risks as using any other vessel for conducting physical inspections. Both are intrusive enough to need express provision in the agreement itself. 
 
Acoustic detection is therefore possible in principle, but its feasibility is heavily dependent on the specific geography and the counterintelligence posture of the monitored state. With that said, the more aggressive the countermeasures a state deploys (e.g.\ restricting fishing vessels, imposing exclusion zones, or deploying acoustic decoys), or the more restrictive a state is over the monitoring regimes a treaty may permit, the more suspicion those measures themselves attract. For example, if a state rejected a treaty clause mandating the bilateral deployment of seabed hydrophone networks or regular AUV surveys (akin to IAEA inspection protocols \citep{simon2025}) this would generate significant suspicions of concealment effort.
 
\subsection{Optical and SAR Imaging Satellites}
Neither optical satellites nor synthetic aperture radar (SAR) can detect an operational UDC directly. Optical sensors require reflected sunlight, which attenuates rapidly in seawater from 30--35\,m depths. SAR uses targeted microwave pulses which, while capable of penetrating cloud cover and operating at night, produce near-zero radar return from submerged objects. While SAR techniques have been used to detect bottom topography and depth gradients via surface-wave modulation \citep{wiehle2019}, a UDC would likely not produce an SAR-detectable perturbation in the local wave field.
 
Optical and SAR imaging are nonetheless relevant as deployment-phase detection tools. As noted earlier, there are approximately 60 ships worldwide capable of laying submarine cables, with only a small subset capable of deploying the power cables required to deliver hundreds of megawatts to a seabed cluster. SAR can be used to identify and classify specific vessel types, which can be fused with satellite Automatic Identification System (AIS) data to flag vessels that have disabled or modified their transponders --- a technique already used in sanctions enforcement and fishing regulation \citep{gfwdark, windward}. Similarly, optical and SAR satellites have been used in open-source-intelligence contexts to monitor shipyard activity, track construction progress, and infer production rates \citep{nti}. Assuming the pool of shipyards capable of manufacturing pressure vessels for UDCs is 20--30 globally, the same methodology would allow monitors to track abnormal production volumes or unscheduled deployments.
 
The key limitation for optical and SAR monitoring is timing: a cable installation completed before a verification regime came into force, or conducted in a manner consistent with other commercial activity, would not generate a distinguishing signature. However, given the maintenance required for frontier training that was outlined earlier, it is likely that a recurring surface signature would be generated and detected throughout a cluster's operational cycle.
 
\subsection{Synthesis}
Taken together, these three modalities suggest that an operational, frontier-scale UDC could potentially be concealed completely; however, it is much more likely that a UDC would be detected during construction or maintenance if SAR and optical satellites are actively used for surveillance of the cable-laying vessel fleet and fabrication yards. Whilst thermal detection is unlikely to be viable for anything other than operational UDCs in very shallow or calm waters, acoustic detection could also be valuable as a verification mechanism depending on the inspection rights an agreement establishes.

\section{Conclusion}\label{sec:conclusion}

Underwater data centers are technically feasible as a dark-compute pathway for 100{,}000 H100-equivalent training runs, but only under demanding and strategically distinctive conditions. The predominant obstacle seems to be the large amount of hands-on maintenance currently required during large AI training runs, which is incredibly difficult underwater. A state actor willing to absorb significant cost and timeline premiums could plausibly overcome this, but the resulting facility is strategically uneconomic --- one that makes sense only if the operator's primary objective is concealment rather than cost or performance efficiency. One important implication is that UDCs built for commercial purposes and UDCs built for evasion have very distinct engineering signatures.
 
No single detection modality forecloses the UDC evasion pathway, but a layered approach raises the bar substantially. Deployment-phase surveillance of the small pressure-vessel fabrication pool and the AIS-tracked power-cable-laying vessel fleet are the most effective detection pathways. Working in conjunction with supply-chain-focused measures, satellite imagery is likely powerful at tracking UDC buildout. Detecting a UDC once it has been built will be much harder, but the maintenance might still be clearly detectable. Small hydrophones and AUVs provide a promising pathway to detection and verification, especially if parties to an agreement mutually agree to host them in areas where UDCs may possibly be built. Overall, the detection methodologies discussed work best in a layered approach, where one layer compensates for the weaknesses of another.
 
How feasible a path to dark compute UDCs present depends in large part on the available alternatives. Underground data centers --- concealed in mines or excavated cavities, with off-grid power and construction disguised as civil or mining activity --- offer substantially greater ease of hardware maintenance. Concealment of data centers as ordinary industrial buildings presents an even lower barrier: equipment can be delivered in ordinary shipments, personnel movements are unremarkable, and the construction signature is minimal compared to the marine logistics operation that any frontier-scale UDC requires. Compared with UDCs, these data centers may be easier to detect after they have been built, but this is likely compensated for by the greater difficulty of hiding UDC construction.
 
Given the obstacles outlined and the feasibility of detection during the buildout stage, we believe that, while in principle possible, UDCs likely constitute a much harder path to treaty evasion compared with the alternatives. Nevertheless, we recommend that the intelligence agencies of nation states actively monitor developments in this area and research detection methodologies.

\appendix
\section{Derivations}\label{app:derivations}
The estimates below support the analysis in Section~\ref{sec:detect}. Each is an order-of-magnitude calculation intended to bound the quantity in question rather than to model it precisely; the assumptions behind each are stated inline.

\subsection{Cluster footprint}\label{app:footprint}
Seabed area required by a frontier-scale cluster.

Order-of-magnitude estimate. Assuming 150--200\,MW required power, and 2\,MW per pod (feasible given demonstrated designs) $\approx$ 75--100 pods.\footnote{Whilst 2 MW per pod exceeds any independently verified underwater deployment, this seemed more realistic for dark-compute on the relevant timescales. At 1 MW per pod the cluster becomes 150–200 pods, yielding a footprint of 3.4–18 ha.} Assume 15--30\,m center-to-center spacing between pods (for deployment logistics and/or heat-discharge flow), i.e.\ 225--900\,m$^2$ total space needed per pod. 75 $\times$ 225\,m$^2$ $\approx$ 1.7\,ha; 100 $\times$ 900\,m$^2$ $\approx$ 9\,ha.

\subsection{Minimum current velocity}\label{app:velocity}
Lowest current velocity consistent with evading thermal detection.

In the most extreme case:
200\,{MW} heat rejected $\div$ 
($4.0\times10^{6}\,\text{J m}^{-3}\text{K}^{-1}$
[seawater heat capacity]
$\times 0.3^{\circ}\text{C}$
[maximum covert SST anomaly]
$\times 100\,\text{m}$
[lower than the densest feasible cluster width]
$\times 10\,\text{m}$
[mixing depth under summer stratification])
$\approx 0.17\,\text{m/s}$ as the lowest current velocity viable for evasion.

\subsection{Temperature rise in calm water}\label{app:temprise}
Surface temperature anomaly in the low-current limit.

In the most conservative case:
$150\,\text{MW} \div$
($4.18\times10^{6}\,\text{J m}^{-3}\text{K}^{-1}$
[freshwater heat capacity]
$\times 0.005\,\text{m/s}$
[wind-driven circulation]
$\times 300\,\text{m}$
[widest feasible cluster width]
$\times 35\,\text{m}$
[full water column mixing])
$\approx 0.68\,^{\circ}\text{C}$ 
(rising to $\sim 2.4\,^{\circ}\text{C}$ under summer stratification, where mixing depth is limited to $10\,\text{m}$).

\subsection{Acoustic source level}\label{app:sourcelevel}
Order-of-magnitude source level for a frontier-scale cluster.

This is an order-of-magnitude estimate ($\pm15\,\text{dB}$). 
The average audible sound levels for liquid-immersed network transformers and step-voltage regulators in the $1{,}001$--$2{,}000\,\text{kVA}$ range are $60$--$61\,\text{dB}$ \citep[Table~2]{nema}.\footnote{We treat the tabulated figure as A-weighted and as an approximate $1\,$m free-field equivalent. Converting a $0.3\,$m contour average to $1\,$m via the sound-power route would add roughly $5\,\text{dB}$, which sits inside the $\pm15\,\text{dB}$ range.} A-weighting attenuates by $19.1\,\text{dB}$ at $100\,\text{Hz}$, so after adding $\sim12$--$19\,\text{dB}$ (with the lower bound allowing for energy distributed at higher harmonics), we get a tone level of $\sim72$--$80\,\text{dB re }20\,\mu\text{Pa}$. 
Converting to an equal-intensity underwater pressure equivalent gives $\sim133$--$142\,\text{dB re }1\,\mu\text{Pa}$, after adding $26\,\text{dB}$ for the pressure-reference change and $\sim35$--$36\,\text{dB}$ for the air/water acoustic-impedance difference.\footnote{Whilst we use the 61.5\,\text{dB} factor for simplicitly, \citep{finfer2008} sets out the limitations of this approach.} 
Subtracting $\sim15$--$25\,\text{dB}$ for structural hull coupling, and adding $18.8$--$20\,\text{dB}$ for incoherent summation across the $75$--$100$ pods ($10\log_{10}75$ to $10\log_{10}100$) yields a source level of $\sim127$--$147\,\text{dB re }1\,\mu\text{Pa}$ at $1\,\text{m}$. 
For reference, the source level of an offshore wind turbine can be approximated as $156\,\text{dB re }1\,\mu\text{Pa}$ at $1\,\text{m}$, with tones below $1\,\text{kHz}$ \citep{tougaard2020}. 

\subsection{Acoustic detection range}\label{app:detrange}
Passive detection range against a coastal ambient-noise floor.

Detection requires the received level (source minus transmission loss) to exceed the ambient-noise level in the transformer's approximate frequency. 
At $\sim100\,\text{Hz}$ a shipping-dominated ambient floor is $75$--$80\,\text{dB re }1\,\mu\text{Pa}^2/\text{Hz}$ \citep[Fig.~1, p.~11]{hildebrand2009};\footnote{Hildebrand's Fig.~1 is a generalised spectrum for a deep-water site with the receiver at $1{,}000\,$m. Coastal ambient noise at this frequency is generally higher and more variable, so adopting it here is a detection-favouring (conservative) choice of floor.} 
integrating the transformer tone over a narrow $\sim1\,\text{Hz}$ band and applying modest narrowband processing gain gives an effective detection threshold of $\sim75\,\text{dB re }1\,\mu\text{Pa}$. 
With a $127$--$147\,\text{dB}$ source, shallow-water three-halves spreading \citep{ainslie2010} at $100\,\text{Hz}$ (TL $\approx 15\log_{10} r$) and negligible absorption, the maximum tolerable transmission loss is $\sim52$--$72\,\text{dB}$, giving a nominal detection range of $\sim3$--$60\,\text{km}$. 
Importantly, three-halves spreading is a mid-range approximation; at ranges of many hundreds of water depths in a $35\,\text{m}$ duct, mode stripping and bottom interaction raise transmission loss above this model, and the achievable range becomes strongly dependent on bottom type and sea state. A working figure of order $5$--$20\,\text{km}$ is more defensible, with the upper tail reserved for highly favourable conditions.

\newpage
\section{Additional Figures}

\begin{figure}[!htbp]
    \centering
    \includegraphics[width=0.5\linewidth]{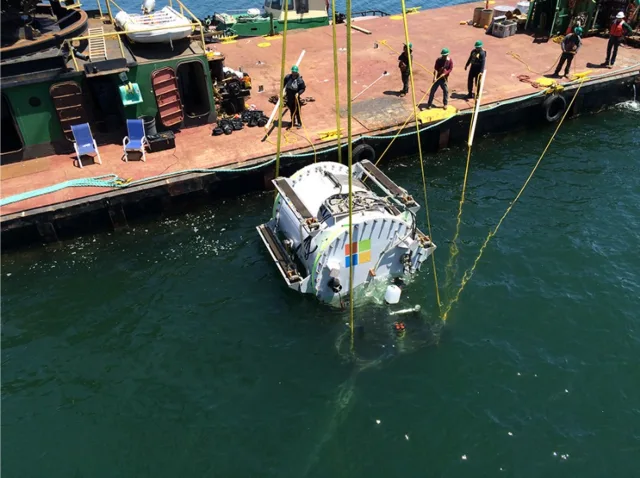}
    \caption{Phase 1 Natick pod is deployed off California's coast. (Image courtesy of Microsoft).}
    \label{fig:Natick1}
    
    \vspace{2pt}
    
    \includegraphics[width=0.5\linewidth]{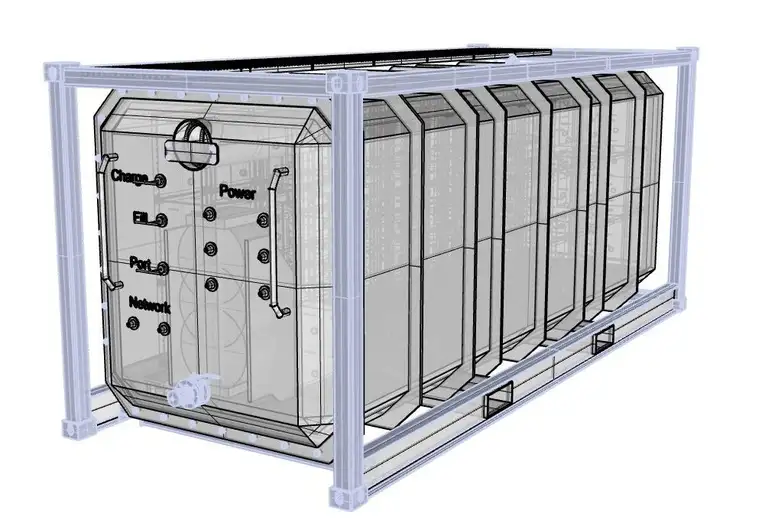}
    \caption{A rendering of Subsea Cloud's Jules Verne pod. (Image courtesy of Subsea Cloud).}
    \label{fig:SubseaVerne}
    
    \vspace{2pt}
    
    \includegraphics[width=0.5\linewidth]{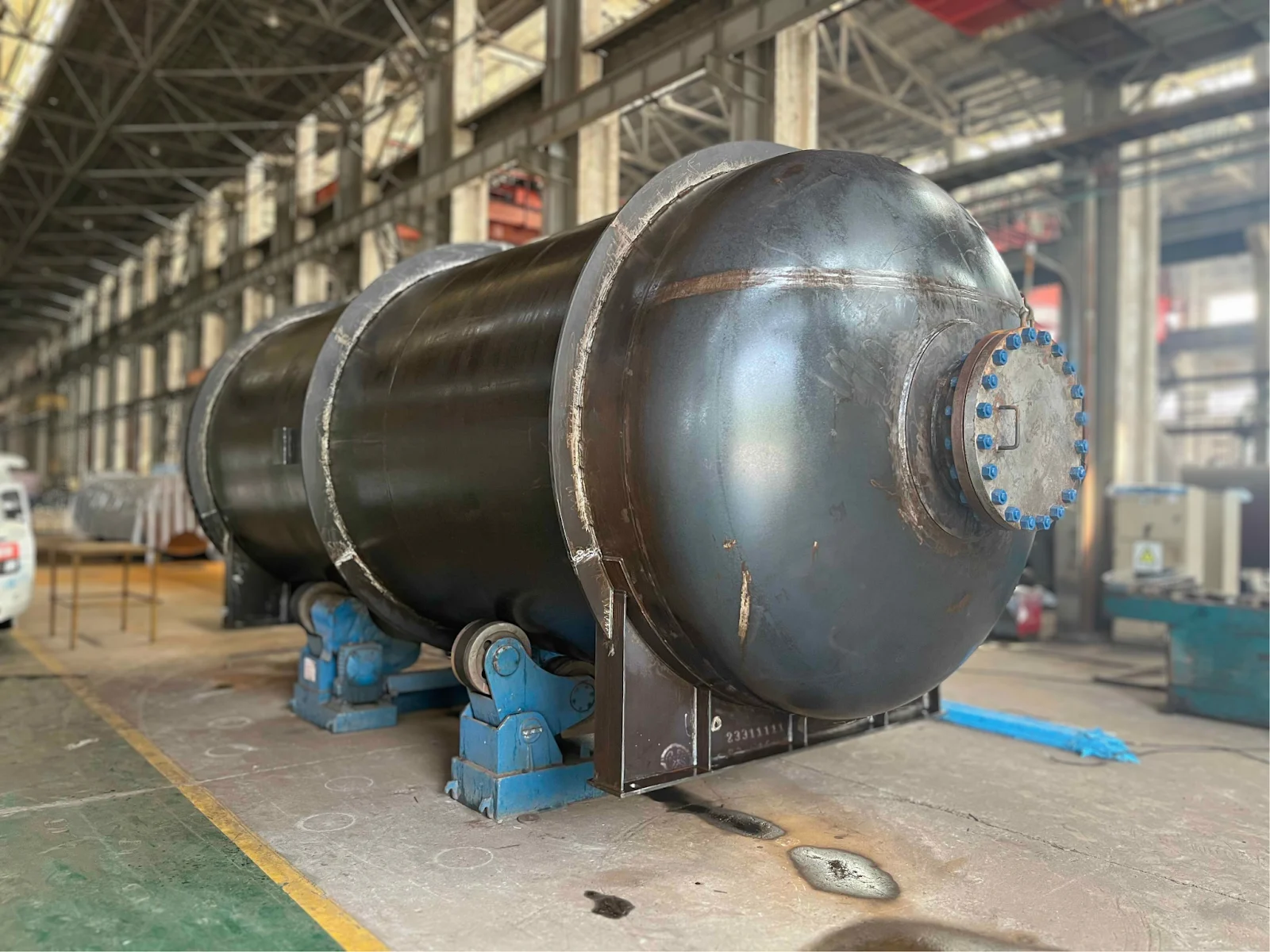}
    \caption{An image of Network Ocean's 0.5\,MW pod under construction. (Image courtesy of Network Ocean).}
    \label{fig:Network}
\end{figure}

\newpage
\bibliography{references} 
 
\end{document}